\documentstyle[12pt]{article}

\newlength{\dinwidth}                                                           
\newlength{\dinmargin}                                                          
\catcode`@=11

\def\@citex[#1]#2{\if@filesw\immediate\write\@auxout{\string\citation{#2}}\fi
  \def\@citea{}\@cite{\@for\@citeb:=#2\do
    {\@citea\def\@citea{,\penalty\@m}\@ifundefined
      {b@\@citeb}{{\bf ?}\@warning
       {Citation `\@citeb' on page \thepage \space undefined}}%
\hbox{\csname b@\@citeb\endcsname}}}{#1}}

\def\citer{\@ifnextchar [{\@tempswatrue\@citexr}{\@tempswafalse\@citexr[]}}

\def\@citexr[#1]#2{\if@filesw\immediate\write\@auxout{\string\citation{#2}}\fi
  \def\@citea{}\@cite{\@for\@citeb:=#2\do
    {\@citea\def\@citea{--\penalty\@m}\@ifundefined
       {b@\@citeb}{{\bf ?}\@warning
       {Citation `\@citeb' on page \thepage \space undefined}}%
\hbox{\csname b@\@citeb\endcsname}}}{#1}}
\catcode`@=12

\def\bo{{\raise.15ex\hbox{\large$\Box$}}}               
\def\face{{\raise.2ex\hbox{$\displaystyle \bigodot$}\mskip-2.2mu \llap {$\ddot
        \smile$}}}                                      

\def\Zbf{{\bf Z}}

\def\leftrightarrowfill{$\mathsurround=0pt \mathord\leftarrow \mkern-6mu
        \cleaders\hbox{$\mkern-2mu \mathord- \mkern-2mu$}\hfill
        \mkern-6mu \mathord\rightarrow$}       
\def\dvec#1{\vbox{\ialign{##\crcr
        \leftrightarrowfill\crcr\noalign{\kern-1pt\nointerlineskip}
        $\hfil\displaystyle{#1}\hfil$\crcr}}}           

\def\beq{\begin{equation}}
\def\eeq{\end{equation}}

\def\beqx{\begin{displaymath}}
\def\eeqx{\end{displaymath}}

\def\beql{\begin{eqnarray}}
\def\eeql{\end{eqnarray}}

\begin{document}	

\begin{flushright}
NIKHEF/00-007\\
April 2000
\end{flushright}

\vspace{15mm}
\begin{center}
{\Large\bf\sc Crosscaps, Boundaries and T-duality \\ }
\end{center}
\vspace{2cm}
\begin{center}
{\large L.R. Huiszoon and  A.N. Schellekens}\\
\vspace{15mm}
{\it NIKHEF Theory Group\\
P.O. Box 41882, 1009 DB Amsterdam, The Netherlands} \\
\end{center}

\vspace{2cm}

\begin{abstract}

Open descendants with boundaries and crosscaps
 of non-trivial automorphism type are studied. We focus on the case where the bulk symmetry is broken to a $Z_2$ orbifold subalgebra. By requiring positivity and integrality for the open sector, we derive a unique crosscap of automorphism type $g \in Z_2$ and a corresponding $g$-twisted Klein bottle for a charge conjugation invariant. As a specific example, we use T-duality to construct the descendants of the true diagonal invariant with symmetry preserving crosscaps and boundaries.  

\end{abstract} 

\thispagestyle{empty}

\newpage \setcounter{page}{2}

\section{Introduction}

The general prescription to construct open unoriented strings from closed oriented ones is known as the method of open descendants~\cite{sagnotti}. It is not limited to circle compactifications and orbifolds of circles, as orientifolds are. In short, one has to find a set of {\em crosscap} and {\em boundary coefficients} from which one can calculate the Klein bottle, annulus and M\"obius strip partition function that together with the torus generate the full spectrum of the open unoriented string. Various consistency conditions~\cite{cardy}~\cite{sewing}~\cite{crosscap}~\cite{nondiagonal}~\cite{completeness}~\cite{zuber} constrain these coefficients. However, most of these constraints require an explicit knowledge of OPE-coefficients and duality (fusing and braiding) matrices, which are only known in a limited number of cases. There is however one very powerful consistency condition that can be applied to all conformal field theories; the partition functions in the open sector have to generate positive and integral state multiplicities. In all known cases~\cite{bert}, this condition determines the crosscap coefficients uniquely once the boundary coefficients are known.

Historically, much progress on open descendants was made after discoveries in conformal field theory on surfaces with boundaries. Cardy~\cite{cardy} derived the boundary coefficients that describe symmetry preserving boundary conditions in case that the closed strings are described by a charge conjugation invariant. Sagnotti and collaborators~\cite{descendants}~\cite{planar} found a formula for the crosscap coefficient, thus completing the open descendants for the `C-diagonal' case. The boundary coefficients for some non-charge conjugation invariants were derived in~\cite{class}. In~\cite{nondia}~\cite{bert} 
we 
constructed open descendants for these theories by deriving the unique crosscap coefficients that, together with the boundary coefficients, satisfy positivity and integrality of the open sector partition function.  

Due to the work of Fuchs and Schweigert~\cite{class}~\cite{DCFT}~\cite{SBB}, the boundary coefficients for boundary conditions that leave an orbifold subalgebra of the full (`bulk') symmetry invariant are now known. In this letter, we will complete the construction of open descendants for this case. The organization of this letter is as follows: In section~\ref{sec-orb}, we briefly review the relation between $Z_2$ orbifolds and simple current extensions. In section~\ref{sec-int}, the open descendants of an integer spin invariant in the orbifold theory are presented. We derive two inequivalent Klein bottles for the integer spin invariant. These results are interpreted in terms of the extended theory in section~\ref{sec-t}. We will show that T-duality appears in an elegant way. As a specific example of our results, we construct the descendants of the true diagonal invariant with symmetry preserving boundary conditions. All technicalities are confined in the appendix.

\section{Orbifolds and simple currents} \label{sec-orb}

In this letter we will consider the following situation. We start with
a theory with a chiral algebra ${\cal A}$ and an order two, integer
spin simple current $J$. We can then extend ${\cal A}$ by $J$, and
obtain a new theory, whose chiral algebra we will denote as ${\cal A}^{\rm E}$.
The fields of the ${\cal A}$-theory
can be divided into three classes: uncharged,
non-fixed fields (labeled $i_0$); charged fields (labeled $i_1$) and
fixed fields (labeled $f$), which are always uncharged. The fields of the
${\cal A}^{\rm E}$ theory are then labeled by the 
uncharged, length two orbits (formed by a pair of fields $i_0$ and $Ji_0$),
plus {\it two} fields for each fixed point $f$. Choosing an orbit 
representative we denote the former as $[i_0]$, and the latter as
$[f,\psi]$, where $\psi$ is a $\Zbf_2$ 
character\rlap.\footnote{Denoting the $\Zbf_2$ elements as $(1,g)$,
the characters are explicitly $\psi_0: \psi_0(1)=\psi_0(g)=1$, and 
$\psi_1: \psi_1(1)=1, \psi_1(g)=-1$. 
We will furthermore use the convention that $\psi$, without
arguments, denotes the value of the character on $g$, which
is precisely what distinguishes the two characters.} 

Conversely,
the chiral algebra ${\cal A}^{\rm E}$ 
has an order two automorphism, denoted $\omega$
and 
there are two kinds
of irreducible highest weight representations~\cite{birke}: {\em symmetric} fields $[i_0]$ are invariant under the action of $\omega$ and {\em non-symmetric} fields $[f,\psi]$ transform as $\omega[f,\psi] = [f,-\psi]$. 
This automorphism has the special property that it leaves
the Virasoro algebra, and in particular the conformal weights fixed. 

Such automorphisms are of interest in open string constructions, since
one may consider boundaries and crosscaps that 
are ``$\omega$-twisted" in the sense of \cite{SBB}. 
Such boundaries or crosscaps do not preserve the full chiral algebra, but
only the sub-algebra ${\cal A}$. For string consistency this must include
the Virasoro algebra and in particular $L_0$, which explains why one must
require that conformal weights are preserved exactly.
A well-known example of such an automorphism is charge conjugation. The symmetric fields are then the self-conjugate fields, and the non-symmetric fields are the complex fields. Another example is the permutation automorphism of a tensor product of two identical conformal field theories~\cite{BHS}.
The point is now that any such automorphism can be described most
easily in terms of the $\omega$-orbifold theory of the 
${\cal A}^{\rm E}$-theory, and that orbifold theory is precisely the 
${\cal A}$-theory\rlap.\footnote{Although our notation is similar to
~\cite{SBB}~\cite{birke}, there is one important difference: 
the chiral algebras ${\cal A}$ and $\bar {\cal A}$ of these authors are
denoted  respectively ${\cal A}^{\rm E}$ and ${\cal A}$ in the present letter.
This is because the ``orbifold" theory plays the most prominent r\^ole in
our work, and we want to avoid excessive occurrences of ``bars" in formulas.}

Our approach is thus to take the orbifold theory ${\cal A}$ as the 
starting point, and extend in the closed sector by the current $J$ to
${\cal A}^{\rm E}$. 
Then we find corresponding boundaries and crosscaps, but without
insisting that they also have the full symmetry ${\cal A}^{\rm E}$. 
From the point of view of the ${\cal A}^{\rm E}$-theory this goes by the name
of ``broken bulk symmetry", whereas from the point of view of the 
${\cal A}$ theory the term ``extended bulk symmetry" would seem more
appropriate.

\section{Open descendants of integer spin invariants} \label{sec-int}

In this section, we will focus on the ${\cal A}$ theory. 
Consider the following modular invariant torus partition function~\cite{simple}\footnote{This invariant is a product of an integer spin invariant and charge conjugation. The charge conjugate of a field $i$ is denoted by $i^c$. } 
\begin{equation} \label{eq:tor}
	Z^J = \sum_{ij} Z^J_{ij} {\chi}_i \bar{\chi}_j = \sum_{i_0, {\rm Rep}} ({\chi}_{i_0} + {\chi}_{Ji_0}) (\bar{\chi}_{i_0^c} + \bar{\chi}_{Ji_0^c})  + 2 \sum_{f} {{\chi}}_{f} {\bar{\chi}}_{f^c} \;\;\;\; ,
\end{equation}
where $i$ denotes a generic field in the orbifold and ${\chi}_{i}$ the corresponding (Virasoro) character. The first sum in the last expression is over all representatives of integer charge orbits.

We will proceed as follows. We first give the boundary coefficients, which
were presented in~\cite{SBB}. 
From these, we will derive two crosscap coefficients. We will find
that one of them satisfies all open and closed string positivity and 
integrality conditions. The other only satisfies those conditions if 
the boundary conditions of~\cite{SBB} are modified.
In the next section, we will interpret these results in terms of the ${\cal A}^{\rm E}$ theory.   

Three sets of labels have to be distinguished in the following. The
transverse channel labels belong to fields propagating in the bulk. They are the fields that according to the torus partition function are paired with their charge conjugate, with a multiplicity given by the order of the untwisted stabilizer. In this case, the latter equals the stabilizer ${\cal S}_m$, 
the group of simple currents that fix $m$. Hence, the transverse channel labels are the chargeless {\em fields} $m$ 
({\it i.e.} $i_0$ or $f$)
with a multiplicity label $\psi_m$ which is the character of ${\cal S}_m$. 
Note that the multiplicity label is 
trivial if $m=i_0$. The boundary labels distinguish different boundaries. They were determined
in ~\cite{SBB} by considering the {\it classifying algebra}. The result is that
the boundary labels are in one-to-one correspondence with the {\em orbits},
but with an extra multiplicity : $[\alpha, \psi_{\alpha}] = [i], [f,\psi]$, where $[i]=[i_0],[i_1]$. The third kind of label that occurs in the
following is simply the primary field label of the ${\cal A}$-theory, which
will be denoted as $i$. The relevant quantities appearing in the positivity
conditions are the boundary coefficients $B_{ma}$, where $m$ is a 
generic transverse channel label $(m, \psi_{m})$ and $a$ a generic boundary label $[\alpha, \psi_{\alpha}]$ and the crosscap coefficients $\Gamma_m$. 
In terms of these,
the direct annulus, M\"obius and Klein bottle are respectively
\begin{equation} \label{eq:directampl} 
 A^{i}_{~ab}=\sum_m S^{i}_{~m} B_{ma} B_{mb}; \ \ \ 
M^{i}_{~a}=\sum_m P^{i}_{~m} B_{ma} \Gamma_{m}; \ \ \
K^{i}=\sum_m S^{i}_{~m} \Gamma_{m} \Gamma_{m} \ ,
\end{equation}
with 
the understanding that $S^{i}_{~[f,\psi]} \equiv S^{i}_{~f}$, since
the ${\cal A}$-characters on which $S$ act do not depend on $\psi$.
Our conventions and normalizations are as in our previous
papers \cite{klein} and \cite{nondia}. In particular, the reflection
coefficients $R_{ma}=B_{ma}\sqrt{S_{m0}}$ satisfy
$\sum_m R_{ma}R_{mb}^*=\delta_{ab}$ and $\sum_a R_{ma}R_{na}^*=\delta_{mn}$.

The boundary coefficients are
\begin{equation} \label{eq:bound} 
	B_{(m, \psi_m)[\alpha, \psi_\alpha]} =  {\sqrt{|\cal G|} \over 
\sqrt{|{\cal S}_{m}|}|{\cal S}_{\alpha}|}{ \sum_J \psi_{m}(J) \psi_{\alpha}(J) {S}^J_{m\alpha} \over \sqrt{{S}_{0m}}} \;\;\;\; ,
\end{equation}
where $|{\cal S}_{m}|$ is the dimension of the stabilizer of $m$,
and $|{\cal G}| = 2$ is the dimension of the simple current group.
 The sum is over all currents in the intersection ${\cal S}_m \cap {\cal S}_\alpha$. The matrix ${S}^0 \equiv {S}$ is the usual S-matrix of the ${\cal A}$ theory and~\footnote{Note that in~\cite{nondia} $\breve{S}$ is defined differently: it is related to $S^J$ by a phase.}  ${S}^J \equiv \breve{S}$ is the fixed point resolution matrix
for the current $J$. 
These matrices are explicitly known for WZW-models~\cite{FSS} and extended WZW-models~\cite{WZWE}. This result (\ref{eq:bound})
is obtained from \cite{SBB}, apart from the normalization,
which we have adapted to our conventions.

The direct annulus can be computed from the boundary coefficients using
(\ref{eq:directampl}):
\begin{eqnarray}
	{A}_{[j][k]}  & = & \sum_i (
{N}^i_{~jk} + 
{N}^{Ji}_{~jk}) 
{\chi}_{i} \;\;\;\; , \\
	{A}_{[j][g,\psi']} & = & \sum_i {N}^i_{~jg} {\chi}_{i} \;\;\;\; , \\
	{A}_{[f,\psi] [g,\psi']} & = & {1\over 2} \sum_i ({N}^i_{~fg} + 
\psi \psi' \breve{N}^{i}_{~fg}) {\chi}_{i} \;\;\;\; , \label{eq:anfixed}
\end{eqnarray}
where ${N}$ are the fusion coefficients of the orbifold theory and \begin{equation}
	\breve{N}^{i}_{fg} = \sum_{m} {\breve{S}_{fm} \breve{S}_{gm} {S}^*_{im} \over {S}_{0m}} \;\;\;\; .
\end{equation}
Note that the annuli have the following property due to (monodromy) charge conservation. When the charges of the boundary labels are equal, only untwisted sector fields contribute to the sums. When the boundary labels have a different charge, i.e., in mixed annuli, only twisted sector fields contribute.
Furthermore all characters appear in ${\cal A}^{\rm E}$ linear
combinations $\chi_i + \chi_{Ji}$, although these combinations are 
${\cal A}^{\rm E}$-characters only if $i$ has zero charge. 

Let us now turn to the crosscap coefficients. They can be derived in a similar way as was done in~\cite{bert}. That is, we have to require that the M\"obius strip satisfies the positivity and integrality relation
\begin{equation} \label{eq:posint}
	 |M^i_{[\alpha, \psi_{\alpha}]}| \leq A^i_{[\alpha, \psi_{\alpha}][\alpha, \psi_{\alpha}]} \;\;\;\;\; {\rm and} \;\;\;\;\; M^i_{[\alpha, \psi_{\alpha}]} = A^i_{[\alpha, \psi_{\alpha}][\alpha, \psi_{\alpha}]} \;\;\;\; {\rm mod} \;\; 2 \;\;\;\; .
\end{equation}
If we choose the special boundary $[0]$ this implies 
$$ M^i_{[0]}=\varepsilon_1 \delta^i_0 + \varepsilon_2 \delta^{Ji}_0 \ , $$
where $\varepsilon_1$ and $\varepsilon_2$ are signs.  Inverting the
relation between $M$ and the crosscap coefficients gives
\begin{equation} 
\sum_{\psi_m} \Gamma_{(m,\psi_m)} = {\sqrt{|{\cal S}_m| \over |{\cal G}|}}
 {\varepsilon_1{P}_{0m} +\varepsilon_2 {P}_{Jm} \over \sqrt{{S}_{0m}}} \;\;\;\; ,
\end{equation}
Note that for fixed points only the sum over $\psi_m$ is determined, not
the coefficients separately. The solution involves therefore a set
of unknown quantities $\delta_m$:
\begin{equation} \label{eq:crosstwo}
 \Gamma_{(m,\psi_m)} = {1\over\sqrt { |{\cal G}||{\cal S}_m|}}
\left(
 {\varepsilon_1{P}_{0m} +\varepsilon_2 {P}_{Jm} \over \sqrt{{S}_{0m}}}
+ \psi_m {\delta_m\over\sqrt{{S}_{0m}}}\right) \;\;\;\; ,
\end{equation}
Intuitively one expects $\delta_m$ to vanish because there is only one
crosscap; indeed, such corrections occur for $B_{ma}$ only for boundary
labels that occur in pairs originating
from a fixed point. We will find that all positivity and integrality
conditions are 
satisfied if $\delta_m=0$. Introducing non-zero $\delta_m$'s 
leads in most cases to violations of these conditions in the closed
and/or open channel or to complex M\"obius coefficients. We can show
that $\delta_m=0$ if $P_{0m}\not=0$ or $P_{Jm}\not=0$ and also that all
$\delta_m$ must vanish if $S^J$ is purely real or imaginary (as it is
in most cases), but we cannot rule out $\delta_m$ in all imaginable cases.
We will therefore assume from now on that it vanishes. Then the
crosscap coefficients are fixed up to two signs 
$\varepsilon_1$ and $\varepsilon_2$.

Requiring positivity and integrality for a boundary label $[f,\psi]$ fixes the relative sign (see appendix), and the overall sign is in any case
never fixed by CFT considerations. Up to this overall sign,
the result is 
\begin{equation} \label{eq:cross}
\Gamma^+_{(m,\psi_m)} = {1 \over\sqrt{|{\cal S}_m| |{\cal G}|}} {{P}_{0m} +\epsilon {P}_{Jm} \over \sqrt{{S}_{0m}}} \;\;\;\; ,
\end{equation}
where $\epsilon \equiv e^{\pi {\rm i}h_J}$ is a sign. The meaning of the superscript `$+$' becomes clear later. 
Now we can compute the direct Klein bottle
\begin{equation} K^{++} = \sum_{i, Q_J(i)=0} ({Y}_{i00} +\epsilon {Y}_{i0J}) {{\chi}}_{i} \;\;\;\; .
\end{equation}
Since this Klein bottle was derived using open sector positivity 
constraints (in fact, just a few of them), it is perhaps 
surprising that it satisfies the positivity and integrality condition for the closed sector (see appendix~\ref{sec-apclos} for details).

Let us assume for the sake of definiteness that ${\cal A}^{\rm E}$ has
complex representations, with conjugation corresponding to
$[f,\psi]^c=[f,-\psi]$. (This amounts to taking $\omega=C$, i.e., charge conjugation. All of the following holds in more general 
situations). As we will explain in the next section, the invariant~(\ref{eq:tor}) can either be interpreted as a charge conjugation invariant or a diagonal invariant for the extension ${\cal A}^{\rm E}$. We will also see that the Klein bottle $K^{++}$ is a standard Klein bottle for the charge conjugation invariant; it projects on world-sheet parity $\Omega$ invariant states. However, $K^{++}$ is a ``twisted Klein bottle" for the diagonal invariant, which means that it projects on $\omega\Omega$ invariant states for this invariant (see subsection~\ref{sec-KBtwist}). Recall that the crosscap we derived is unique~\footnote{A possible non-uniqueness of the crosscap due to the $\delta_m$ in the crosscap coefficient~(\ref{eq:crosstwo}) cannot provide us a standard Klein bottle for the diagonal invariant: in case of $\omega=C$, $S^J$ is purely imaginary so all $\delta_m$ must vanish.}. So in order to find a standard Klein bottle projection for the diagonal invariant, we either have to change the boundary coefficients or the positivity and integrality condition of the open sector. Suppose for the moment that we keep the boundary conditions~(\ref{eq:bound}) fixed. Instead of~(\ref{eq:posint}), we require a ``$\omega$-twisted" positivity and integrality condition
\begin{equation} \label{eq:posint2}
	 |M^{i}_{[\alpha, \psi_{\alpha}]}| \leq A^i_{[\alpha, \psi_{\alpha}][\alpha, -\psi_{\alpha}]} \;\;\;\;\; {\rm and} \;\;\;\;\; M^i_{[\alpha, \psi_{\alpha}]} = A^i_{[\alpha, \psi_{\alpha}][\alpha, -\psi_{\alpha}]} \;\;\;\; {\rm mod} \;\; 2 \;\;\;\; .
\end{equation}
It is easy to see that we can now derive a unique crosscap given by
\begin{equation} \label{eq:cross2}
\Gamma^-_{(m,\psi_m)} = {1\over\sqrt{|{\cal S}_m| |{\cal G}|}}{{P}_{0m} -\epsilon {P}_{Jm} \over \sqrt{{S}_{0m}}} 
\end{equation}
and corresponding Klein bottle 
\begin{equation} K^{--} = \sum_{i, Q_J(i)=0} ({Y}_{i00} -\epsilon {Y}_{i0J}) {{\chi}}_{i} \;\;\; ,
\end{equation}
which is inequivalent to $K^{++}$ but also satisfies positivity and integrality of the closed sector. This Klein bottle has the desired property that it is a standard Klein bottle for the diagonal invariant (see subsection~\ref{sec-KBtwist}).   

Alternatively, one may leave the positivity and integrality conditions~(\ref{eq:posint}) unchanged, but modify the boundary conditions simply by
replacing $S^J$ by ${\rm i}S^J$ in (\ref{eq:bound}). This flips the sign in the annulus~(\ref{eq:anfixed}) for two fixed point boundary labels and leads straightforwardly to the crosscap~(\ref{eq:cross2}). This is reminiscent of what happens if one chooses different Klein bottle
projections in the Cardy case, as in \cite{descendants}\cite{klein}. If
one leaves the boundary coefficients unchanged, one may
encounter contributions like ${1\over2}({\cal N}_a^2+{\cal N}_b^2)\chi_0$ in the 
open string partition function, where $\chi_0$ is the identity character and ${\cal N}$ the CP factors.
Changing the appropriate boundary conditions by a factor ${\rm i}$ changes this
to  ${\cal N}_a{\cal N}_b\chi_0$. For ${\cal N}_a={\cal N}_b$ the latter can be interpreted in term
of a $U({\cal N}_a)$ gauge group, whereas the former (even though for ${\cal N}_a={\cal N}_b$ it
is numerically equal) does not seem to allow a gauge group interpretation.
Therefore we think changing the boundary coefficients is the correct
interpretation. It is not clear to us whether this affects the analysis
of \cite{SBB}, in which (\ref{eq:bound}) is derived from
the sewing constraint for the  bulk-bulk-boundary correlator.

Finally, we display the M\"obius strip amplitudes. In the
direct channel (open string loop channel) they are
\begin{eqnarray} \label{eq:M1}
	{M}^{\pm}_{[j]} & = & \sum_{i} ({Y}_{j0}^{~~i} \pm \epsilon {Y}_{jJ}^{~~i}) \hat{{\chi}}_{i} \;\;\;\; ,\\ \label{eq:M2}
	{M}^{\pm}_{[f,\psi]} & = & {1 \over 2} \sum_{i} ({Y}_{f0}^{~~i} \pm \epsilon {Y}_{fJ}^{~~i}) \hat{{\chi}}_{i}\;\;\;\; ,
\end{eqnarray}
where the $\pm$ refer to the crosscap (~\ref{eq:cross} or~\ref{eq:cross2}) that appear in the transverse M\"obius strip. By (monodromy) charge conservation and equation~(\ref{eq:yn}) of the appendix, only chargeless fields contribute. In the appendix, we show that $M^+$ and $M^-$ satisfy the positivity and integrality conditions (\ref{eq:posint}) and (\ref{eq:posint2}) respectively for all boundary labels.   

One is tempted to consider, 
as the notation might suggest, the introduction of `Chan-Paton factors' ${\cal M}_{\pm}$ for the crosscap coefficients $\Gamma^{\pm}$ in the same way as normal CP-factors are introduced for boundary labels. However, it is not hard to show that only twisted states ($Q(i) \not = 0$) contribute to ``mixed Klein bottles" $K^{+-}$. 
Since these states do not occur in the torus partition function,
the requirement of positivity and integrality of the closed sector, ie, equation~(\ref{eq:posK}), forces us to put one of the ``crosscap CP-factors" to zero. 
From now on, we will switch to a more economical notation by defining $K^{++} \equiv K^+$ and $K^{--} \equiv K^-$. 

Note that the boundary and crosscap coefficients that appeared in this section are very similar to the coefficients of the descendants of order two half-integer spin invariants~\cite{nondia}. In that case however, the two different sets of crosscap coefficients have a different origin; they correspond to simple current Klein bottles~\cite{klein}.

\section{Open descendants and T-duality}\label{sec-t}

As already stressed in section~\ref{sec-orb}, the orbifold theory ${\cal{A}}$ has an integer spin simple current $J$ by which we can extend the chiral algebra. The result of this extension is simply ${\cal{A}}^{\rm E}$. The characters of the ${\cal{A}}^{\rm E}$ theory are related to those of the orbifold as follows:
\begin{equation} \label{eq:char}
	\chi_{[i_0]} = {\chi}_{i_0} + {\chi}_{Ji_0} \;\;\;\;\;\;\; \chi_{[f,\psi]} = \chi_{[f,-\psi]} = {\chi}_f \;\;\;\; .
\end{equation}
The invariant (\ref{eq:tor}) can therefore be interpreted as a charge conjugation invariant $Z^c$ of the ${\cal{A}}^{\rm E}$ theory {\em or} as an invariant that is the product of charge conjugation and $\omega$, denoted by $Z^{c\omega}$:
\begin{equation}
	Z^c = \sum_{IJ} C_{IJ} \chi_I \chi_J \;\;\;,\;\;\; Z^{c\omega} = \sum_{IJ} C_{I,\omega J} \chi_I \chi_J \;\;\;\; ,
\end{equation}
where $I,J$ are generic fields of the ${\cal{A}}^{\rm E}$ theory. These invariants are known to be T-dual. By T-duality, we simply mean a ``one-sided $\omega$ transformation" that acts on closed string states as
\begin{equation}
	{\rm T} : |I,J \rangle \rightarrow |I,\omega J \rangle \;\;\;\; ,
\end{equation}   
which is a duality for every automorphism $\omega$ that preserves the conformal weights exactly.

Note that $T$-duality acts not only on the ground states but also on the
currents in  ${\cal A}^{\rm E}/{\cal A}$.
This implies in particular that the definition of the Ishibashi~\cite{cardy}~\cite{DCFT}~\cite{RS}
states flips: ${\cal A}^{\rm E}$-symmetric Ishibashi states are turned into
Ishibashi states of automorphism type~\cite{DCFT}~\cite{RS} $\omega$ and vice-versa. This is 
a direct consequence of the fact that in one of the chiral algebras
$J$ is replaced by $\omega(J)$.

{}From the point of view of the ${\cal A}$ theory T-duality is trivial, since
the automorphism $\omega$ is defined only after resolution of the fixed
points. Hence all boundary and crosscap states and all amplitudes are
invariant under T-duality. T-Duality becomes non-trivial only once
we interpret the result in terms of the ${\cal A}^{\rm E}$-theory.

A second clue to the automorphism type of crosscaps and boundaries is
the expression one obtains in terms of ${\cal A}$ characters for the
direct Klein bottle, annulus and M\"obius strip. If these amplitudes 
are obtained from two ${\cal A}^{\rm E}$
symmetry-preserving boundaries/crosscaps, they must be expressible in terms of
${\cal A}^{\rm E}$ characters; conversely if they cannot be expressed
in terms of ${\cal A}^{\rm E}$ characters, at least one of 
the boundary/crosscap states must be symmetry breaking. Since the amplitudes
are T-duality invariant, but the automorphism type flips, it follows that
also amplitudes obtained with two boundaries/crosscap of automorphism
type $\omega$ must be expressible in terms of ${\cal A}^{\rm E}$
characters, since this is the case in the T-dual theory.
To summarize, an amplitude can be written in terms of characters of the extension if and only if the two boundaries/crosscaps are of the same automorphism type. 

In the previous section we observed
that in the ``mixed" annuli only twisted sector fields ($Q(i)\not=0$) 
contribute. On the other hand, both Klein bottles $K^+$ and $K^-$ as well
the other annuli can be expressed in terms of ${\cal A}^{\rm E}$ characters.
It is not hard to show that also the ``mixed" M\"obius strips $M^-_{[i_0]}$, $M^-_{[f,\psi]}$ and $M^+_{[i_1]}$ cannot be written in terms of characters of the extension. 

\subsection{Twisted Klein bottles}~\label{sec-KBtwist}

Before we discuss T-duality for the Klein bottles, we have to introduce $g$-{\em twisted Klein bottles}. When $g$ is a symmetry of a closed string theory that commutes with world-sheet parity $\Omega$, we can divide out~\cite{ORIENT} the group $(1,g\Omega)$. At the level of partition functions, we have to add a (halved) $g$-twisted Klein bottle to the (halved) torus. Only eigenstates of $g\Omega$ appear in the $g$-twisted Klein bottle, that is, the Klein bottle coefficients satisfy a $g$-twisted positivity and integrality condition 
\begin{equation} \label{eq:twistK}
	|K_{I}| = Z_{I,g I} 
\end{equation}
with the modular invariant $Z$. What kind of Klein bottles did we find in the last section? From the appendix~\ref{sec-apclos}, we know
\begin{equation}
	|K^{+}_I| = C_{II} \;\;\;,\;\;\; | K^{-}_I| = C_{I,\omega I} \;\;\;\; .\end{equation}
When we compare with equation~(\ref{eq:twistK}), we come to the following conclusion: $K^+$ is an untwisted Klein bottle from the point of view of the charge conjugation invariant $Z^{c}$ and a $\omega$-twisted Klein bottle for the invariant $Z^{c\omega}$. For $K^-$ it is the other way around: it is an $\omega$-twisted Klein bottle from the point of view of the charge conjugation invariant and an untwisted Klein bottle for the invariant $Z^{c\omega}$. Note that the operations $\omega \Omega$ and $\Omega$ are T-dual. 

\subsection{Symmetry breaking crosscaps}

We have observed above
that $K^+$ is the standard Klein bottle for the charge conjugation invariant. Indeed, from equation~(\ref{eq:p}) of the appendix, it follows that the corresponding crosscap coefficient is just that of the `C-diagonal case' $\Gamma^+_I = P_{0I}/\sqrt{S_{0I}}$. This coefficient is generically non-vanishing for all $I$, so the corresponding automorphism type $g$ has to satisfy $Z_{I,gI^c}=1$ for all $I$ . Therefore $g=1$ and thus
$\Gamma^+$ must have trivial automorphism type from the point of view of the charge conjugation invariant $Z^c$. Hence it has
automorphism type $\omega$ for the invariant $Z^{c\omega}$. 

Since only one of the M\"obius amplitudes $M^+$ and $M^-$ for a given boundary label can be written in terms of characters of the ${\cal A}^{\rm E}$ theory, it follows that $\Gamma^+$ and $\Gamma^-$ must have opposite automorphism types. So $\Gamma^-_I$ is trivial for $Z^{c\omega}$ and of automorphism type $\omega$ for the invariant $Z^{c}$. 
An important check on this interpretation is the fact that 
the ${\cal A}^{\rm E}$-Ishibashi states $[f,\psi]$ are 
not present in the theory if one uses the $Z^c$ modular invariant.
Hence the ${\cal A}^{\rm E}$ crosscap should vanish for the 
corresponding transverse channel labels. 
Indeed, one can show that $\Gamma^-_I$ vanishes identically for fixed points.

\subsection{Symmetry breaking boundaries}

Given the symmetry properties of the crosscaps and those of the M\"obius
strip, one can now read off those of the boundary coefficients.
In agreement with ~\cite{SBB} we find that
from the point of view of a charge conjugation invariant $Z^c$, the boundary coefficients~(\ref{eq:bound}) have the following automorphism types. When the charge of the boundary label is zero, the boundaries leave the ${\cal{A}}^{\rm E}$ algebra  invariant; they are of trivial automorphism type. Charged boundaries $[i_1]$ are of automorphism type $\omega$; they only leave the orbifold subalgebra ${\cal A}$ invariant. From the point of view of $Z^{c\omega}$, the automorphism types of the boundary conditions are reversed: the chargeless boundaries break the symmetry, whereas the charged boundaries do not. This is of course nothing but a reformulation of the well-known fact~\cite{ORIENT} that Dirichlet (automorphism type $\omega$) and Neumann (trivial automorphism) boundary conditions are interchanged under T-duality. 
A similar check can be made as in the last paragraph of the previous section.
If we use the $Z^{c\omega}$ modular invariant, the symmetric boundary
coefficients $B_{(f,\psi)[i_1]}$ should vanish if $f$ is a fixed point. This is
indeed true, as a consequence of the fact that $S$ vanishes between fixed
points and charged fields.

\subsection{Open descendants of diagonal invariants}

Let us first conclude: there are two types of inequivalent open descendants for the charge conjugation invariant $Z^c$. In the first one, we project on $\Omega$ invariant states with the Klein bottle $K^+$. Via the channel transformation, this leads to crosscaps that preserve the full bulk symmetry. The open sector has two kinds of boundary conditions; chargeless boundary conditions have trivial automorphism and charged boundaries have automorphism type $\omega$. A second descendant can be constructed when we project on $\omega\Omega$ invariant states with a $\omega$-twisted Klein bottle $K^-$ that satisfies a twisted positivity and integrality condition. The corresponding crosscaps have automorphism type $\omega$. The open sector has again two kinds of boundary conditions; chargeless boundary conditions with trivial automorphism and charged boundaries with automorphism type $\omega$. In order to satisfy positivity and integrality of the open sector, some boundary coefficients differ by a factor $i$ relative to those of the untwisted Klein bottle projection as explained in the previous sector.

We could equally well have started with the invariant $Z^{c\omega}$. The automorphism types of boundaries and crosscaps are opposite to those of the charge conjugation invariant. T-duality relates both invariants and also the corresponding open descendants. 

As a specific example, take $\omega = C$, i.e., charge conjugation. We can now construct the open descendants of a diagonal invariant $Z_{IJ} = \delta_{IJ}$ with crosscaps and boundaries of trivial automorphism type. By T-duality, this is equivalent to a charge conjugation invariant with crosscaps and boundaries of automorphism type $C$. So we have to take the Klein bottle $K^-$ and put the CP-factors of the chargeless boundaries to zero. The standard Cardy case~\cite{cardy}~\cite{descendants}, i.e., a charge conjugation invariant with trivial crosscaps and boundaries, can be obtained in a similar way: take $K^+$ and put the CP-factors of the charged boundaries to zero. 

\vspace{10mm}
\begin{center}
{\bf Acknowledgements}
\end{center}
We would like to thank N. Sousa, J. Fuchs and C. Schweigert
for useful discussions. L.H. would like to thank the 
``Samenwerkingsverband Mathematische Fysica" for financial support.
\vspace{10mm}

\appendix

\section{Positivity and integrality}

Let us first relate the fusion and Y-fusion coefficients of the ${\cal A}^{\rm E}$ theory to those of the orbifold ${\cal A}$.  Let us first be a bit more general, and allow the simple current group to be $\cal G$. We denote a generic field of the ${\cal A}^{\rm E}$ theory by $[{i}, \psi_{i}]$. Fields in the ${\cal A}$ theory are denoted by $i, j$. We will not add the superscript ${\rm E}$ to quantities of the ${\cal A}^{\rm E}$ theory, since the indices attached to these quantities make the formulas unambiguous. The S-matrix of the extension is given by~\cite{FSS}
\begin{equation}
	 S_{[{i}, \psi_{i}][{j}, \psi_{j}]} = {|{\cal G}| \over |{\cal S}_{i}||{\cal S}_{j}|} \sum \psi_{i}(J) \psi_{j}(J)^* {S}^J_{{i}{j}} \;\;\;\; .
\end{equation}
This gives the fusion coefficients via the Verlinde formula
\begin{equation} \label{eq:fus}
	{N}^{\quad\quad\quad\quad[{j}_3, \psi_{j_3}]}_{[{j_1}, \psi_{j_1}][{j_2}, \psi_{j_2}]} = \sum_{[m]} \sum_{\psi_m} {{S}_{[m, \psi_m][{j_1}, \psi_{j_1}]} {S}_{[m, \psi_m][{j_2}, \psi_{j_2}]} {S}^*_{[m,\psi_m] [{j_3}, \psi_{j_3}]} \over {S}_{[m,\psi_m][0]}} \;\;\;\; .
\end{equation}
The P-matrix of the extension is~\cite{nondia}
\begin{equation} \label{eq:p}
	P_{[{i}, \psi_{i}][{j}, \psi_{j}]} = {|{\cal G}| \over |{\cal S}_{i}||{\cal S}_{j}|} \sum_J \psi_{i}(J) \psi_{j}(J)^* \hat{P}^J_{{i}{j}} \;\;\;\; ,
\end{equation}
where the sum is over the intersection ${\cal S}_{i} \cap {\cal S}_{j}$ and where 
\begin{equation}
	 \hat{P}^J_{{i}{j}} = {1 \over |{\cal G}|} \sum_K e^{\pi i[h_i - h_{Ki}]} {P}^J_{Ki,j} \;\;\;\; ,
\end{equation}
and where the sum is now over {\em all} $K$ and where
\begin{equation}
	 {P}^J = \sqrt{{T}}{S}^J{T}^2{S}^J\sqrt{{T}} \;\;\;\; .
\end{equation}
The Y-fusion coefficients are given by~\cite{planar}
\begin{equation} \label{eq:pfus}
	{Y}^{\quad\quad\quad\quad[{j}_3, \psi_{j_3}]}_{[{j_1}, \psi_{j_1}][{j_2}, \psi_{j_2}]} = \sum_{[m]} \sum_{\psi_m} {{S}_{[m, \psi_m][{j_1}, \psi_{j_1}]} {P}_{[m, \psi_m][{j_2}, \psi_{j_2}]} {P}^*_{[m,\psi_m] [{j_3}, \psi_{j_3}]} \over {S}_{[m,\psi_m][0]}} \;\;\;\; .
\end{equation}

Recall~\cite{bantay}~\cite{klein} that $Y_{i00}$ is the Frobenius-Schur indicator of a field $i$ in a conformal field theory; its value is (minus) one for (pseudo) real fields and zero for complex fields. Furthermore, the tensor $Y$ is integral and satisfies a ``positivity and integrality relation" with the fusion coefficients~\cite{bantay}~\cite{BHS}~\cite{klein}:
\begin{equation} \label{eq:yn}
	 |Y_{i0}^{~~j}| \leq N_{ii}^{~~j} \;\;\;, \;\;
 Y_{i0}^{~~j} = N_{ii}^{~~j} \;\;\; {\rm mod} \;\; 2 \;\;\;\; ,
\end{equation}
which plays a crucial r\^ole in all proofs that follow.

\subsection{Positivity and integrality of the closed sector} \label{sec-apclos}

In this subsection, we prove that the Klein bottle coefficients, given by 
\begin{equation} \label{eq:klein}
	K^\pm_i = {Y}_{i00} \pm \epsilon {Y}_{i0J} \;\;\;\;\;\; , \;\;\;\; \epsilon = e^{\pi i h_J } \;\;\;\; ,
\end{equation}	
satisfy 
\begin{equation} \label{eq:posK}
	|K^{\pm}_i| \leq Z^J_{ii} \;\;\;, \;\; K^{\pm}_i = Z^J_{ii} \;\;\; {\rm mod} \;\; 2 \;\;\;\; .
\end{equation}
From the torus~(\ref{eq:tor}) we see that there are three kinds of fields that appear on the diagonal: self-conjugate fixed points $f=f^c$, self-conjugate $i_0=i_0^c$ and fields that satisfy $i_0 = Ji_0^c$. We first concentrate on the fixed points. From~(\ref{eq:fus}) and~(\ref{eq:pfus}) we find 
\begin{equation}
 	{N}_{[0][f,\psi][f,\pm\psi]} = {1\over 2} ({N}_{0ff} \pm \breve{N}_{0ff}) \;\;\;\; , \;\;\;\;\;  Y_{[f,\psi][0][0]} = {1\over 2} ({Y}_{f00} +\epsilon {Y}_{f0J}) \;\;\;\; .
\end{equation}
We can distinguish three situations:
\begin{itemize}
	\item $[f,\psi]^c = [f,\psi]$. So $Y_{[f,\psi][0][0]} = \pm 1$ which implies ${Y}_{f00} = \epsilon {Y}_{f0J} = \pm 1$. The corresponding Klein bottles~(\ref{eq:klein}) therefore satisfy 
\begin{equation}
	|K^{+}_f| = 2 \;\;\;\;\; , \;\;\;\;\;  |K^{-}_f| = 0 \;\;\;\; ,
\end{equation}
which satisfies~(\ref{eq:posK}) since $f=f^c$. 
	\item $[f,\psi]^c = [f,-\psi]$. So $N_{[0][f,\psi] [f,-\psi]} = 1$ which implies ${N}_{0ff} = - \breve{N}_{0ff} = 1$ and $f$ is self-conjugate. Furthermore, since $Y_{[f,\psi][0][0]} = 0$ we have ${Y}_{f00} = -\epsilon {Y}_{f0J} = \pm 1$. The corresponding Klein bottles~(\ref{eq:klein}) therefore satisfy 
\begin{equation}
	|K^{+}_f| = 0 \;\;\;\;\; , \;\;\;\;\;  |K^{-}_f| = 2 \;\;\;\; ,
\end{equation}
which satisfies~(\ref{eq:posK}) as well.
	\item $[f,\psi]^c \neq [f,\psi]$ and $[f,\psi]^c \neq [f,-\psi]$. In this case $f \neq f^c$ and both Klein bottles vanish in agreement with~(\ref{eq:posK}). 
\end{itemize}
Now we turn to the fields $i_0$. Equations~(\ref{eq:fus}) and~(\ref{eq:pfus}) give 
\begin{equation}
 	{N}_{[0][i_0][i_0]} = {N}_{0i_0i_0} + {N}_{Ji_0i_0} \;\;\;\; , \;\;\;\;\;  Y_{[i_0][0][0]} = {Y}_{i_000} +\epsilon {Y}_{i_00J} \;\;\;\; .
\end{equation}
There are now two different cases:
\begin{itemize}
	\item $[i_0]^c = [i_0]$. So ${N}_{[0][i_0][i_0]} =1$ and either $i_0 = i_0^c$ or $i_0 = Ji_0^c$. When $i_0^c=i_0$, ${Y}_{i_000} =\pm 1$ and ${Y}_{f0J} = 0$ and when $i_0 = Ji_0^c$ it is the other way around. In any case
\begin{equation}
	|K^{\pm}_{i_0}| = 1 \;\;\;\; .
\end{equation}
	\item $[i_0]^c \neq [i_0]$. So ${N}_{[0][i_0][i_0]} =0$ which implies $i_0 \neq i_0^c$ and $i_0 \neq Ji_0^c$. Both Klein bottles vanish for these fields, in agreement with~(\ref{eq:posK}).
\end{itemize}
So the Klein bottle of section~\ref{sec-int} satisfies positivity and integrality. In section~\ref{sec-t}, we regard the Klein bottles as projections for the theory described by ${\cal A}^{\rm E}$. Note that we have to be careful in case of fixed points. Since one fixed point of the orbifold theory resolves into two fields $[f,\psi]$ of the extension, the same happens for the corresponding Klein bottle coefficients; the coefficient $K_f = \pm 2$ splits into two coefficients $K_{[f,\psi]} =\pm 1$ and $K_{[f,-\psi]} =\pm 1$.  We will assume that $[f,\psi]$ and $[f,-\psi]$ have the same Klein bottle coefficient, so that a coefficient $K_f = 0$ in the orbifold theory cannot split in a $K_{[f,\psi]} = 1$ and $K_{[f,-\psi]} = -1$ for instance. This is required by the {\em Klein bottle constraint}~\cite{descendants}~\cite{klein}, which forbids $[f,\psi]$ and $[f,-\psi]$ to have opposite Klein bottle coefficients when $[f,\psi]^c = [f,-\psi]$, a situation that occurs generically. 

{}From the above analysis, it follows that the Klein bottles satisfy
\begin{equation}
	|K^{+}_I| = C_{II} \;\;\;\;\; , \;\;\;\;\; |K^{-}_I| = C_{I,\omega I} \;\;\;\; ,
\end{equation}
where $I$ is a generic field in the ${\cal A}^{\rm E}$ theory.

\subsection{Positivity and integrality of the open sector} \label{sec-apopen}

In this section, we prove that the two pairs of M\"obius and annulus coefficients from section~\ref{sec-int} satisfy 
\begin{equation} \label{eq:MA}
	 |M^{\pm}_{[\alpha, \psi_{\alpha}]i}| \leq A^\pm_{[\alpha, \psi_{\alpha}][\alpha, \psi_{\alpha}]i} \;\;\;\;\; {\rm and} \;\;\;\;\; M^\pm_{[\alpha, \psi_{\alpha}]i} = A^\pm_{[\alpha, \psi_{\alpha}][\alpha, \psi_{\alpha}]i} \;\;\;\; {\rm mod} \;\; 2 \;\;\;\; ,
\end{equation}
for all boundary labels $[\alpha, \psi_{\alpha}]$ and all fields $i$. The annuli $A^\pm$ are not defined explicitely in the main text. By $A^+$ we denote the annulus that corresponds to the boundary coefficient~(\ref{eq:bound}) and by $A^-$ the annulus of the modified boundary coefficient. It differs from $A^+$ by a relative minus sign when both boundary labels are fixed points.  

For the boundary labels $[i] = [i_0], [i_1]$, equation~(\ref{eq:MA}) follows immediately from equation~(\ref{eq:yn}). For the fixed point boundary labels, we have to prove
\begin{eqnarray} \label{eq:PI}
	{1 \over 2} |{Y}_{f0i} \pm \epsilon {Y}_{fJi}| & \leq & {1 \over 2} ({N}_{iff} \pm \breve{N}_{iff}) \;\;\;\; ,\\ {1 \over 2} ({Y}_{f0i} \pm \epsilon {Y}_{fJi}) & = & {1 \over 2} (N_{iff} \pm \breve{N}_{iff}) \;\;\; {\rm mod} \;\; 2 \;\;\;\; ,
\end{eqnarray}
where $i=i_0,g$. In order to prove these relations, it is convenient to do a similar trick as was done in~\cite{nondia}. So we first tensor the $\cal A$ theory with a theory $\bar{\cal A}$ that has an order two integer spin simple current $\bar{J}$ and fixed points $\bar{f}$. Let us denote the fields of the tensor theory ${\cal A}^t$ by $I=(i,\bar{i})$. The ($Y$-)fusion rules of this theory are simply 
\begin{equation} \label{eq:ten}
	N^t_{(i,\bar{i})(j,\bar{j})(k,\bar{k})}= N_{ijk}\bar{N}_{\bar{i}\bar{j}\bar{k}} \;\;\;\;,\;\;\; Y^t_{(i,\bar{i})(j,\bar{j})(k,\bar{k})}= Y_{ijk}\bar{Y}_{\bar{i}\bar{j}\bar{k}}\;\;\;\;\; .
\end{equation}
The tensor theory has an order two integer spin simple current $(J,\bar{J})$ by which we can extend to a theory with chiral algebra ${\cal A}^{{\rm E},t}$. The fields in this extension are the chargeless orbits of the tensor theory, denoted by $[I_0] = [(i,\bar{i}) + (Ji,\bar{J}\bar{i})]$ with $Q_J(i)=Q_{\bar{J}}(\bar{i})$ and the resolved fixed points $[f,\psi]$. The ($Y$-)fusion rules for this extension are related to those of the tensor theory as in equation~(\ref{eq:fus}) and~(\ref{eq:pfus}). With the use of~(\ref{eq:ten}), we can then relate the coefficients of ${\cal A}^{{\rm E},t}$ and $\cal A$. Consider
\begin{equation}
	|Y^{{\rm E},t}_{[f,\psi][0][I]}| \leq   N^{{\rm E},t}_{[f,\psi][f,\psi][I]} \;\;\; , \;\;\; Y^{{\rm E},t}_{[f,\psi][0][I]} =   N^{{\rm E},t}_{[f,\psi][f,\psi][I]} \;\;\;\;\;\; {\rm mod} \; 2 \;\;\; ,
\end{equation}
which holds by equation~(\ref{eq:yn}). In this equation, $F=(f,\bar{f})$ and $I=(i,\bar{0})$, where $i=i_0,g$. In terms of quantities of the $\cal{A}$ and $\bar{\cal{A}}$ theory, the above conditions become
\begin{eqnarray}
	{1 \over 2} |Y_{f0i} \bar{Y}_{\bar{f}\bar{0}\bar{0}} + \epsilon_{J,\bar{J}} {Y}_{fJi} \bar{Y}_{\bar{f}\bar{J}\bar{0}}| & \leq & {1 \over 2} (N_{ffi} \bar{N}_{\bar{f}\bar{f}\bar{0}} + \breve{N}_{ffi}\breve{\bar{N}}_{\bar{f}\bar{f}\bar{0}}) \\
	{1 \over 2} (Y_{f0i} \bar{Y}_{\bar{f}\bar{0}\bar{0}} + \epsilon_{J,\bar{J}} {Y}_{fJi} \bar{Y}_{\bar{f}\bar{J}\bar{0}}) & = & {1 \over 2} (N_{ffi} \bar{N}_{\bar{f}\bar{f}\bar{0}} + \breve{N}_{ffi}\breve{\bar{N}}_{\bar{f}\bar{f}\bar{0}}) \;\;\;\;\;\; {\rm mod} \; 2 \;\;\; ,
\end{eqnarray}
where $\epsilon_{J,\bar{J}} = e^{\pi i[h_J + h_{\bar{J}}]}$. Now we use for the $\bar{A}$ theory $B_2$ level $2k$. This theory has an order two, spin $k$ simple current and a fixed point with $\bar{N}_{\bar{f}\bar{f}\bar{0}}=1$ and $\bar{Y}_{\bar{f}\bar{0}\bar{0}} = \bar{Y}_{\bar{f}\bar{0}\bar{J}} = \breve{\bar{N}}_{\bar{f}\bar{f}\bar{0}} = (-1)^k$. So equation~(\ref{eq:PI}) with the (plus) minus sign follows by taking $k$ (even) odd.

\end{document}